# Modeling of Kundu-Eckhaus equation

Dmitry Levko, Alexander Volkov

**Abstract:** In work the numerical solutions of Kundu-Eckhaus equation are presented. The conditions of dominate of all terms are cleared up.

In [1] was shown that Nonlinear Schrödinger equation

$$iq_t + q_{xx} - 2\sigma|q|^2 q = 0 \tag{1}$$

with special gauge transformation

$$q \to Q = q \cdot \exp(i\theta), \tag{2}$$

$$\theta = \delta \int^x dx' \cdot |Q(x')|^2, \tag{3}$$

proceed to Kundu-Eckhaus equation:

$$iQ_t + Q_{xx} - 2\sigma|Q|^2 Q + \delta^2 |Q|^4 Q + 2i\delta(|Q|^2)_x Q = 0. \tag{4}$$

Nonlinear Schrödinger equation (NLSE) (1) appears in different physical contexts: in nonlinear optics [2]-[5], in plasma's physics [6], in fiber optics [7], in Bose-Einstein condensates [8]-[9] and others. It consists soliton solutions [10], [11]. The stability of NLSE solutions is associated with a balance between linear dispersive and nonlinear collapsing terms. But in some physical situations are demanded addition of higher nonlinear terms to (1) (see ref. [12]). Then balance lost and equation doesn't allow analytical solutions. In [1] are shown that NLSE with addition terms (higher order nonlinear terms compensates higher order linear terms) allows analytical solutions.

In this letter we study by numerically methods the solutions of (4).

Introduce phase time $z = x - Vt$, $p = V/2$ and solution of (4) present in such form [13]

$$Q(x,t) = y(z)\exp(i(pz + \omega t)).$$

Then (4) we can write in next form

$$y_{zz} + E_0 y - 2\sigma|y|^2 y + \delta^2 |y|^4 y + 2i \cdot \delta \cdot \left(y^* y' + y(y^*)'\right) y = 0. \tag{5}$$

Here $y(z)$ is complex function. It is related to complex unit before the last term in (5). We can split $y(z)$ on real and imaginary parts

$$y(z) = v(z) + iu(z). \tag{6}$$

Then we find the system of equations

$$\begin{cases} v_{zz} + E_0 v - 2\sigma(v^2 + u^2)v + \delta^2(v^2 + u^2)^2 v - 4\delta(v'v + u'u) \cdot v = 0, \\ u_{zz} + E_0 u - 2\sigma(v^2 + u^2)u + \delta^2(v^2 + u^2)^2 u + 4\delta(v'v + u'u) \cdot u = 0, \end{cases} \tag{7}$$

with initial conditions

$$v(0) = u(0) = 5 \cdot 10^{-5},$$

$$v'(0) = u'(0) = 0.$$

Here $E_0 = p^2 - \omega$. This method of solution KE-equation doesn't allow us to distinguish solitons and solitary waves.

The soliton solutions corresponds the case $E_0 < 0$ and σ<0. It is demonstrated on fig.1-4.

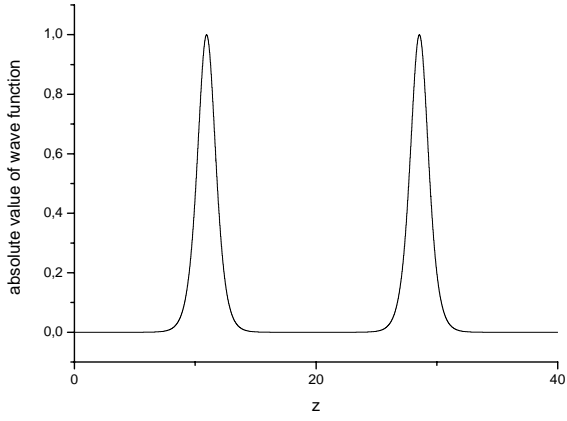

Figure 1. $E_0=-1$, $\sigma=-1$, $\delta=10$

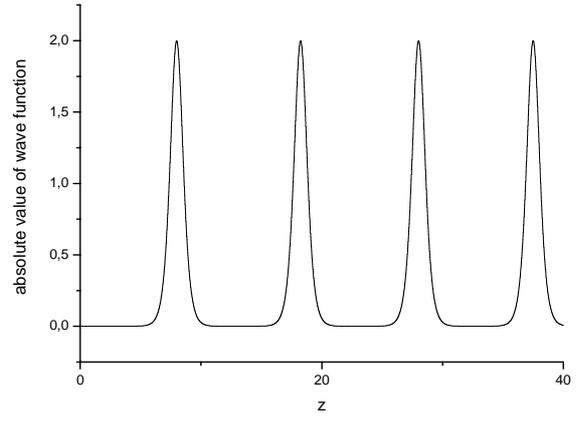

Figure 2. $E_0=-2$, $\sigma=-1$, $\delta=10$

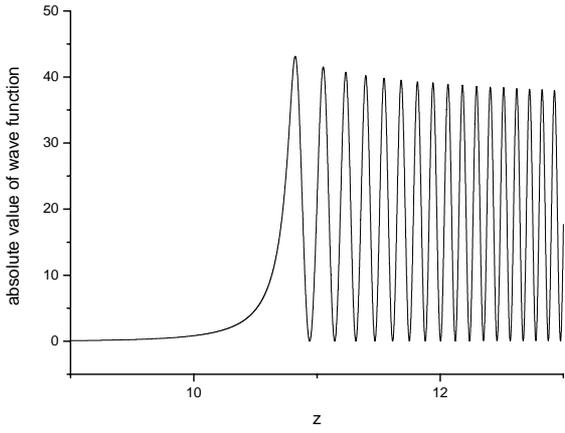

Figure 3. $E_0=-1$, $\sigma=1$, $\delta=10$

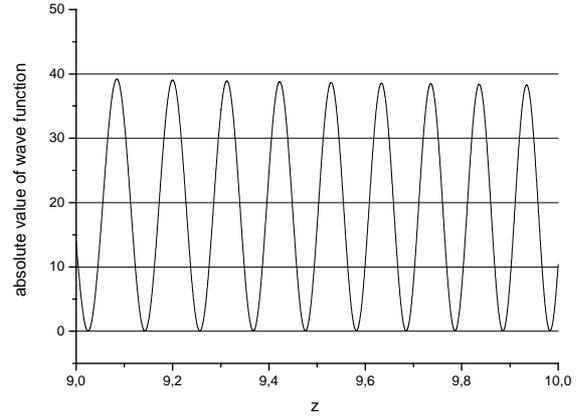

Figure 4. $E_0=-2$, $\sigma=1$, $\delta=10$

From fig.3-4 see that in defocusing case ($\sigma=1$), at first, solution increase to some value and after decrease to some constant level and oscillate. The oscillations are demonstrated on fig.5. Also from fig.1-2 see that if $E_0$ increase then amplitude of solutions also increases and the number of solitons on the same phase time increase.

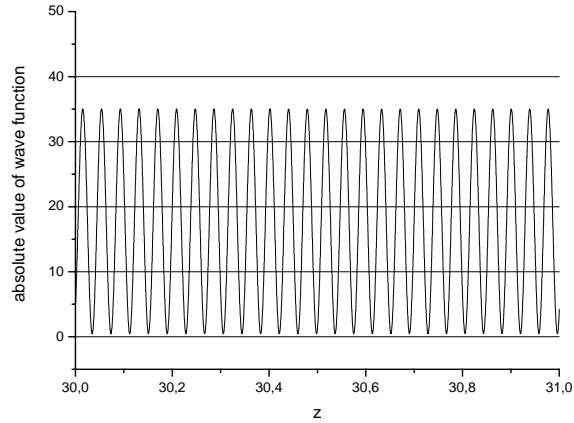

Figure 5. $E_0=-2$, $\sigma=1$, $\delta=10$ (long distances)

The sign of $\sigma$ don't influence on solutions form. It can be explained by domination other terms.
Next eq.(5) present in form

$$y_{zz} + E_0 y - 2\sigma|y|^2 y + a_3\delta^2|y|^4 y + 2i \cdot a_4\delta \cdot \left(y^* y' + y(y^*)'\right)y = 0. \qquad (8)$$

In such form we can study the influence of all terms on the solutions (the introduction the same multiplier before the second and the third terms not essential because it only change the sign of $E_0$ and $\sigma$, respectively). In fig.6-7 the results of solution for $\sigma=\pm 1$ are presented. From fig.6 see that if $a_4$ increase the amplitude of solutions also increase.

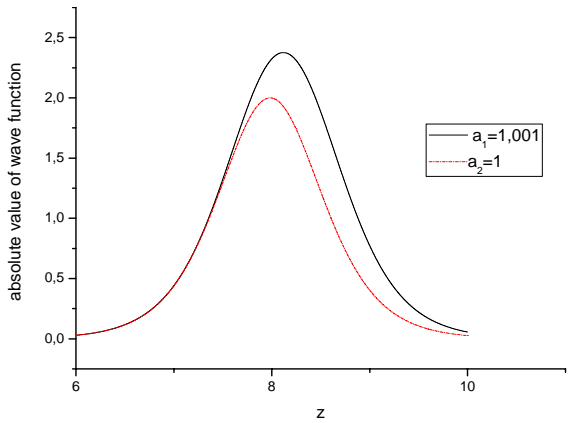
Figure 6. $E_0=-2$, $\sigma=-1$, $\delta=10$

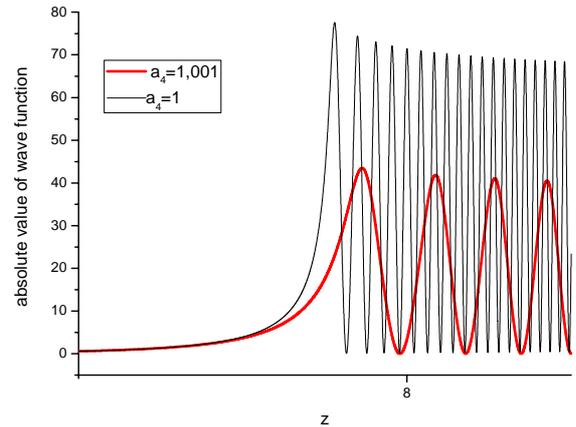
Figure 7. $E_0=-2$, $\sigma=1$, $\delta=10$

From fig.7 see that in the case $\sigma=1$ the behavior of amplitude other than in the case $\sigma=-1$. See that if $a_4$ increase the amplitude decrease.

The influence of the fourth term in (8) is demonstrated on fig.8. See that with increase $a_3$ the amplitude of solutions decrease.

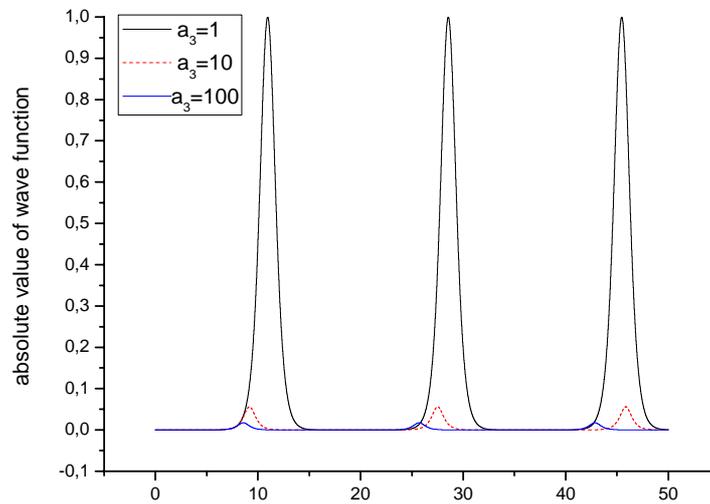
Figure 8. $E_0=-1$, $\sigma=-1$, $\delta=10$

**Literature**

[1] A. Kundu. Symmetry, Integrability and Geometry: Methods and Applications. **Vol.2**, paper 078 (2006).
[2] D. E. Pelinovsky et al. Phys. Rev. E **70**, 036618 (2004).
[3] D. N. Neshev et al. Phys. Rev. Lett. **92**, 123903 (2004).
[4] J. W. Fleischer et al. Phys. Rev. Lett. **92**, 123904 (2004).
[5] Ya. V. Kartashov et al. Phys. Rev. Lett. **96**, 073901 (2006).
[6] A. Kuznetsov et al. Plasma's Physics. **27**, pp.225-234 (2001) (in Russian).
[7] D. Trager et al. arXiv:physics/0601037 v1 7 Jan 2006
[8] L.P. Pitaevskii. Uspekhi Fizicheskikh Nauk **176** (4) 345-364 (2006) (in Russian).
[9] M. Matuszewski et al. arXiv:nlin.PS/0609031 v1 12 Sep 2006.
[10] A. C. Newell. Solitons in mathematics and physics. Mir, Moscow, 1989 (in Russian).
[11] D. Levko. arXiv.org: nlin.SI / 0612027
[12] T. Tao et al. arXiv.org: math.AP / 0511070
[13] A. Frenkin. Moskow Univ. Bull. №2, p.9-12 (1999)

Dmitry Levko[1], Alexander Volkov: Institute of Physics National Ukrainian Academy of Sciences
1. *unitedlevko@yandex.ru*